\documentclass[12pt,tightenlines,showpacs,preprintnumbers,superscriptaddress,nofootinbib]{revtex4}
\pdfoutput=1

\usepackage[pdftex]{graphicx}

\usepackage{slashed}
\usepackage{epsfig}
\newcommand{\PRE}[1]{{#1}} 

\newcommand{\M}{{\cal M}}

\def\beq{\begin{eqnarray}}
\def\eeq{\end{eqnarray}}
\def\bea{\begin{eqnarray}}
\def\eea{\end{eqnarray}}
\def\be{\begin{equation}}
\def\ee{\end{equation}}

\newcommand{\cm}{\text{cm}}

\newcommand{\s}{\text{s}}

\newcommand{\eqref}[1]{Eq.~(\ref{#1})}

\newcommand{\gsim}{\lower.7ex\hbox{$\;\stackrel{\textstyle>}{\sim}\;$}}
\newcommand{\lsim}{\lower.7ex\hbox{$\;\stackrel{\textstyle<}{\sim}\;$}}

\newcommand{\svgg}{\langle\sigma v\rangle_{\gamma\gamma}}
\newcommand{\svgZ}{\langle\sigma v\rangle_{\gamma Z}}
\newcommand{\svgll}{\langle\sigma v\rangle_{\gamma ll}}

\newcommand{\Mdm}{M_{DM}}
\begin{document}
\preprint{ULB-TH/14-10}
\preprint{LPT-Orsay-14-28}



\title{ \PRE{\vspace*{1.5in}} Bremsstrahlung and Gamma Ray Lines\\
 in 3 Scenarios of Dark Matter Annihilation
\PRE{\vspace*{0.3in}} }


\author{Federica Giacchino\PRE{\vspace*{.2in}}}
\affiliation{Service de Physique Th\'eorique\\
 Universit\'e Libre de Bruxelles\\ 
Boulevard du Triomphe, CP225, 1050 Brussels, Belgium\PRE{\vspace*{.2in}}}

\author{Laura Lopez-Honorez}
\affiliation{Theoretische Natuurkunde\\
Vrije Universiteit Brussel and The International Solvay Institutes\\
Pleinlaan 2, B-1050 Brussels, Belgium\PRE{\vspace*{.2in}}}

\author{Michel H.G. Tytgat}
\email{federica.giacchino@ulb.ac.be, llopezho@vub.ac.be, mtytgat@ulb.ac.be}
\affiliation{Service de Physique Th\'eorique\\
 Universit\'e Libre de Bruxelles\\ 
Boulevard du Triomphe, CP225, 1050 Brussels, Belgium\PRE{\vspace*{.2in}}}
\affiliation{Laboratoire de Physique Th\'eorique\\ Universit\'e de Paris-Sud, F-91405 Orsay, France}


\begin{abstract}
Gamma ray spectral features are of interest for indirect searches of
dark matter (DM).  Following Barger {\em et al} we consider 3 simple
scenarios of DM that annihilates into Standard Model (SM) fermion
pairs. Scenario $1$ is a Majorana DM candidate coupled to a charged
scalar, scenario $2$ is a Majorana DM coupled to a charged gauge boson
and scenario $3$ is a real scalar DM coupled a charged vector-like
fermion. As shown by Barger {\em et al}, these 3 scenarios share
precisely the same internal Bremsstrahlung spectral signature into
gamma rays. Their phenomenology is however distinct. In particular for
annihilation into light SM fermions, in the chiral limit, the 2-body
annihilation cross section is p-wave suppressed for the Majorana
candidates while it is d-wave suppressed for the real scalar. In the
present work we study the annihilation into 2 gammas, showing that
these three scenarios have distinct, and so potentially
distinguishable, spectral signatures into gamma rays. In the case of
the real scalar candidate we provide a new calculation of the
amplitude for annihilation into 2 gammas.
\end{abstract}

 \maketitle

\newpage

\section{Introduction}

Dark Matter (DM), which accounts for about 80 \% of all mass in the
universe, is one of the strong indications for physics beyond the
Standard Model (SM) of particle physics. The dominant paradigm is that
dark matter is made of new, neutral and stable (or very long-lived
particles). The most studied possibility is the neutralino, which is
the archetype of a weakly interacting massive particle (WIMP). A WIMP
is a particularly attractive DM candidate.  If the WIMP was in thermal
equilibrium in the early universe, its relic abundance is elegantly
fixed by its annihilation cross section, the matching with
cosmological observations requiring $\langle \sigma
v\rangle~\sim~10^{-26} \cm^3\cdot \s^{-1}$. Also the WIMP hypothesis
may be tested at colliders, using low background detectors (direct
detection) or through the annihilation of DM into SM particles
(indirect detection). One of the issues with indirect searches is that
a potential DM signal may be (and, unfortunately, is expected to be)
obscured by an overwhelming astrophysical background. Hence the
importance of possible so-called smoking gun signatures, {\em i.e.}
signals that have no (or little) astrophysical counterparts, like a
strong gamma-ray line or, more generally, one or many peaks in the
gamma ray spectral energy density (which are called spectral features)
\cite{Bergstrom:1988fp,Rudaz:1989ij,Bergstrom:1989jr} (see
also~\cite{Bringmann:2012ez} for a recent review). Gamma ray features
are actively being searched by the Fermi satellite and the HESS
telescope in the GeV to multi-TeV range. Remarkably, the current constraints
on the annihilation cross section of DM into gamma ray lines are
rather strong, ranging from $\langle \sigma v\rangle~\lesssim~10^{-28}
\cm^3\cdot \s^{-1}$ for $\Mdm \sim 10$ GeV, \cite{Ackermann:2013uma}
to $\langle \sigma v\rangle~\sim~10^{-26} \cm^3\cdot \s^{-1}$ for
$\Mdm \sim 10$ TeV~\cite{Abramowski:2013ax}.

A WIMP is neutral and thus its annihilation in gamma rays lines
proceeds through radiative corrections. In general the annihilation
cross section is suppressed by (powers of) the fine structure constant
$\alpha$ compared to the leading, say, $2 \rightarrow 2$ or 2-body
tree level processes. A notable exception occurs if the 2-body
processes, while being relevant in the early universe, are themselves
suppressed in astrophysical environments, like at the center of our
galaxy. This is for instance possible if the annihilation cross
section is velocity dependent. A familiar example is the annihilation
of a pair of Majorana particles into SM model fermion pairs, in which
case the cross section is mass suppressed and is p-wave in the chiral
limit $\sigma v \propto v^2$~\cite{Goldberg:1983nd}. Another example,
which has been put forward very recently, is annihilation of a real
scalar, again into light fermions, which may be d-wave in the chiral
limit, $\sigma v \propto
v^4$~\cite{Toma:2013bka,Giacchino:2013bta}. In both cases, a simple
consequence is that Bremsstrahlung emission is relatively enhanced,
possibly leading to observable features in the gamma ray spectrum as
well as non negligible contribution at the time of freeze-out
\cite{Bergstrom:1988fp,Baltz:2002we,Bergstrom:2004cy,Toma:2013bka,Giacchino:2013bta}.

Bremsstrahlung of gamma rays and $W$ and $Z$ electroweak gauge bosons
have been extensively studied in the literature, both for their own
sake and with phenomenological applications in mind, see for
instance~\cite{Beacom:2004pe,Boehm:2006df,Bringmann:2007nk,Ciafaloni:2011sa,Bell:2011if,Barger:2011jg,Garny:2011ii,Weiler:2013hh,DeSimone:2013gj,Kopp:2014tsa}. Of
particular interest for the present contribution, Barger {\em et al}
have compared the Bremsstrahlung spectral energy density in three
simple DM scenarios in \cite{Barger:2011jg}. All three scenarios
involve a new charged particle (the mediator) that is chirally coupled
to the DM particle and to SM fermions (which may be leptons or a
quarks). The DM is assumed to be its own antiparticle. For instability
it is also assumed to be odd under some $Z_2$ symmetry, and so is the
charged mediator. The latter must be clearly heavier than the DM
particle. Concentrating on spin $0$ and $1/2$ DM candidates, there are
then three possible scenarios. In scenario $1$, a Majorana DM is
coupled to the SM fermions through a charged scalar. This is similar
to the neutralino, in which case the charged scalar is a slepton or a
squark. In scenario $2$, the DM is also a Majorana particle, but now
it couples to a charged gauged boson. This is possible in some variant
on the Left-Right model \cite{Ma:2009tc}, in which case the DM is some
sort of heavy Majorana neutrino. Finally, in scenario $3$, the DM
is a real scalar coupled to SM fermions through heavy, vector-like
charged fermions. This scenario, which has been developed for other
phenomenological purposes, has been dubbed the Vector-Like Portal in
\cite{Perez:2013nra} (see also \cite{Frandsen:2013bfa} for an
alternative appellation).

In the present article we complement the work of Barger {\em et al}
\cite{Barger:2011jg} and the work we have initiated in
\cite{Giacchino:2013bta}. Concretely, Barger {\em et al} have shown
that, in all three scenarios sketched above, the Bremsstrahlung
spectral signature is precisely the same, up to a normalization that
is scenario dependent. There are good reasons for this, which we
briefly discuss in the next section. In \cite{Giacchino:2013bta} (see
also \cite{Toma:2013bka} in which precisely the same conclusions have
been reached), we have shown that the 2-body annihilation of the
scalar DM candidate is d-wave suppressed in the chiral limit, and
furthermore, that the Bremsstrahlung signal is parametrically
larger. These two effects combined imply that that scenario $3$ may
lead to more significant gamma ray features than a Majorana particle
(specifically scenario $1$). In the same work we had also tentatively
incorporated the contributions of gamma ray lines to the spectral
signatures. In the present work, we compare all 3 scenarios, and in
particular provide analytical expressions for the annihilation of the
DM candidates into 2 gamma rays. In scenario $1$, the result is
well-known and has been derived many times in the literature
\cite{Rudaz:1989ij,Bergstrom:1989jr,Giudice:1989kc,Bergstrom:1997fh},
with which, having redone the calculation, we agree. In scenario $2$,
an analytical expression for annihilation cross section may be
extracted from the results of \cite{Bergstrom:1997fh} in the MSSM. For
lack of time, we do not provide a fully independent check of this
expression. It may be of interest to do so, but having reached the
same result as \cite{Bergstrom:1997fh} in scenario $1$, we have no
reason to doubt their results. In scenario $3$, the full expression is
not available in the literature, so we give it in the present work.
The amplitude is given in~\cite{Bertone:2009cb} in the chiral limit,
and has been used as such e.g. in \cite{Tulin:2012uq} for
phenomenological purposes, but we believe that the result reported
there is incorrect\footnote{We believe that the error, which
  propagated in the literature, is actually just due to a misprint in
  equation (13) of \cite{Bertone:2009cb}, see Sec.~\ref{sec:Sgg}. It
  is however virtually impossible to spot it without knowledge of the
  correct answer. }.  An expression for large mediator mass limit is
also available in~\cite{Boehm:2006gu}.

The plan is as follows. In the next section we begin with a
presentation of the basic features of the three scenarios of
\cite{Barger:2011jg}, including the tree level 2-body annihilation
cross sections and the expressions of the Bremsstrahlung (for emission
of a gamma).  Next we give some details on the one loop calculations
of the DM annihilation into two gamma. In the final section, we
compare the spectral signatures of the three scenarios, and then draw
some conclusions.


\section{3 simple scenarios}
\label{sec:3scenarios}

The 3 scenarios that we consider, following \cite{Barger:2011jg}, are
very simple. They have in common the fact that DM annihilates into SM
fermions through a charged mediator in the t and u channels (we
consider the case of self-conjugate DM candidates). For simplicity we
assume that those channels are the only ones that are relevant, {\em
  i.e.} that other interactions that a given DM candidate may have can
be neglected in some appropriate range of parameters. Hence the
results we discuss may correspond to a corner of all the possible
outcomes of more sophisticated models (for instance scenario $1$
is contained in the MSSM).

\subsection{Scenario $1$:  Majorana DM candidate $\chi$ and  charged scalar $\tilde E$}

The couplings with SM fermions take the form 
\begin{equation}
  {\cal L} \supset y_\chi \tilde E^\dagger \bar \chi P_R \psi_l + h.c.\, .
\label{eq:majDM}
\end{equation}
with $P_R = (1+\gamma_5)/2$.
Although the notation suggests that $\chi$ is coupled only to
right-handed SM leptons $\psi_l$, and so that $\tilde E$ carries a fermionic
charge, the results apply to couplings with quarks, or to $SU(2)$
doublets (modulo more degrees of freedom). Which to choose depends on
the underlying model. Clearly the collider constraints on the mass of
heavy charged fermions (scalars or others) are weaker than those on
particles that carry colour but on the other hand interactions like
that of (\ref{eq:majDM}) are constrained by non-observation of lepton
flavour violating processes, so one may have to compromise (see for
instance \cite{Kopp:2014tsa}). As usually, stability may be simply
insured by imposing a discrete symmetry,
$$
\chi \rightarrow - \chi \quad \mbox{and} \quad \tilde E \rightarrow - \tilde E
$$

In the chiral limit, $m_l \rightarrow 0$, and in the non-relativistic
limit $v_\chi \rightarrow 0$, the 2-body annihilation cross section,
$\chi \chi \rightarrow \bar l l$ is given by
 \begin{equation}
   \sigma v (\chi \chi \rightarrow l\bar l) = {y_\chi^4 \over 48 \pi}\, {v^2\over M_\chi^2}\, {1+ r_\chi^4\over (1+r_\chi^2)^4}
\label{eq:Maj2bdy}
 \end{equation}
where 
$$
r_\chi = {M_{\tilde E}\over M_\chi} \geq 1.
$$ ($v$ is as usual the M\o ller velocity, $v = 2 v_\chi$ in the
center of mass frame \cite{Gondolo:1990dk}).

That the annihilation cross section is p-wave in the chiral limit is
well known \cite{Goldberg:1983nd} and may be stated as follows. A pair
of non-relativistic Majorana DM particles in a s-wave corresponds the
state $^1S_0(O^{-+})$ in the $^{2 S+1}L_J (J^{CP})$ spectroscopic
notation, which, in terms of bi-linear operators, is represented by
$\bar \chi\gamma_5 \chi$. Correspondingly, in a CP conserving theory,
the final state fermion pair is represented by the operator $\bar
\psi_l \gamma_5 \psi_l$, which involves a chirality flip, and is thus
mass suppressed. In a p-wave, the state is $^3 P_1 (1^{++})$, or $\bar
\chi \gamma_k \gamma_5 \chi$, which is coupled to the fermion pair
current, $\bar \psi_l \gamma^kP_R \psi_l$. Hence in the chiral limit,
the annihilation cross section is p-wave.

\subsection{Scenario $2$:  Majorana DM candidate $N$ and  charged gauge boson $W^\prime$}

In this case the mediator is a charged gauge boson, which we call
$W^\prime$. This scenario is akin to the models proposed by Ma {\em et
  al} in \cite{Khalil:2009nb,Ma:2009tc} based on a Left-Right model,
in which the charged gauge boson that couples to right handed (RH)
current carries a generalized fermion number. In that model, unlike
the conventional LR models, the RH neutrino, which we write $N$, is
not the mass partner of the $\nu_L$, but is a viable Majorana DM
candidate. For our purpose we write the coupling of $N$ to
$W^{\prime}$ as
\begin{equation}
{\cal L} \supset {g_N\over \sqrt{2}} W_\mu^{\prime +} \bar N \gamma^\mu P_R \psi_l + h.c.
\label{eq:LN}
\end{equation}
Notice that we have included a factor of $1/\sqrt{2}$, like in the SM,
so our convention for the gauge coupling is different from that of
\cite{Barger:2011jg}.

While the tree level processes may be calculated in a unitary gauge,
the one-loop annihilation cross section that we will rely on has been
calculated in a 't Hooft-Feynman version ($\xi = 0$) of a non-linear
$R_\xi$ gauge (for some details on such gauges, see
\cite{Bergstrom:1994mg}). For this, we also need the coupling of the
$N$ to the nonphysical Goldstone charged scalars, $G^\prime$, which,
one may check, must be given by
\begin{equation}
{\cal L} \supset   {g_N \over \sqrt{2} M_{W^\prime}} G^{\prime +} \bar N\left( M_N P_R - m_l P_L\right)\psi_l\,,
\end{equation}
with  $P_L=(1-\gamma_5)$
.
Although it for sure exists somewhere, we have not found the 2-body
cross section $N N\rightarrow ~\bar l l$ in the literature, so we give
it here, again in the chiral limit $m_f \rightarrow 0$,
\begin{equation}
   \sigma v (N N \rightarrow \bar l l) = {g_N^4 \over 192 \pi}\, {v^2\over M_N^2}\, {(1+ 4 r_N^2 + 13 r_N^4 + 12 r_N^6 + 4 r_N^8)\over r_N^4(1+r_N^2)^4}
\label{eq:Maj2bdy}
 \end{equation}
where now
\begin{equation}
r_N = {M_{W^\prime}\over M_N} \geq 1.
\end{equation}
The dependence on $r_N$ is a bit complicated, but notice that for
large $r_N$ we simply have
\begin{equation}
\label{eq:NN}
\langle\sigma v\rangle (N N \rightarrow \bar l l) \approx \langle v^2\rangle \, {g_N^4 \over 48 \pi}\, { M_N^2\over M_{W^\prime}^4}
 \end{equation}
 the same as for $\chi$
\begin{equation}
\label{eq:chichi}
\langle \sigma v\rangle (\chi \chi \rightarrow \bar l l) \approx \langle v^2\rangle \, {y_\chi^4 \over 48 \pi}\, { M_\chi^2\over M_{\tilde E}^4}
\end{equation}
Of course the 2-body cross section is p-wave for precisely the same
reason as in scenario~$1$.

\subsection{Scenario $3$:  Real scalar  DM candidate $S$ and  charged vector-like fermion  $E$}

In this last scenario, DM is a real scalar particle, $S$ with Yukawa
couplings to a charged vector-like $E$ fermion and the SM fermions
(again we consider couplings to SM singlets for simplicity)
\begin{equation}
  \label{eq:yuk1}
{\cal L} \supset y_S\; S\; \bar E P_R \psi_l + h.c.\,.
\end{equation}
with as above
$$
S \longrightarrow - S$$
and 
$$ E\longrightarrow - E$$ under some discrete $Z_2$ symmetry.
Following~\cite{Perez:2013nra,Giacchino:2013bta} we call this scenario
the Vector-Like Portal.  Being a scalar singlet, $S$ has also a
renormalizable coupling to the SM
scalar~\cite{Silveira:1985rk,Veltman:1989vw,McDonald:1993ex,Burgess:2000yq}.
\begin{equation}
  {\cal L} \supset {\lambda_S\over 2} S^2 \vert H \vert^2\,.
\label{eq:ls}
\end{equation}
We assume that this coupling, if present, is sub-dominant. 

An interesting point about this scenario is that the annihilation
cross section in SM fermions is d-wave in the chiral
limit~\cite{Toma:2013bka,Giacchino:2013bta},
\begin{equation}
\sigma v(S S \rightarrow \bar l l) = { y_S^4 \over 60 \pi}{v^4\over M_S^2}{1\over (1+r^2)^4}
\label{eq:sv2S}
\end{equation}
The suppression by a factor $ v^4$ is a bit unusual but is easy to
understand.  A pair of non-relativistic real scalar DM particles in a
s-wave have quantum numbers $^1 S_0(O^{++})$, corresponding to the
bi-linear operator $S^2$, which may be coupled to SM fermions through
$\bar\psi_l \psi_l$. Hence the amplitude for s-wave annihilation is
mass suppressed, $\propto m_l$. For a S pair, the p-wave state is
$^1 P_1(1^{-+})$ to which corresponds no fermion bi-linear (in a CP
conserving setup)\footnote{If CP is not conserved, or if the S is
  taken to be complex, the state $^1 P_1(1^{--})$ is possible,
  $S^\dagger \partial_k S$, which may be coupled to $\bar \psi_l
  \gamma^k \psi_l$.}. The next possibility is then a d-wave, with $^1
D_2(2^{++})$. This $J=2$ state may be coupled to to SM fermions
through their stress-energy tensor $\Theta^{ij}_l = {i\over 2} \bar
\psi_l (\gamma^i \partial^j - \gamma^j \partial^i) \psi_l$. Hence the
amplitude is d-wave in the chiral limit.

The $v^4$ behaviour has interesting phenomenological implications. In
the early universe one
has~\cite{Gondolo:1990dk,Toma:2013bka,Giacchino:2013bta},
\begin{equation}
  \langle v^2 \rangle = {6 \over x_f}\approx 0.24 \quad \mbox{and} \quad \langle v^4 \rangle = {60 \over x_f^2}\approx 0.1
\label{eq:vfo}
\end{equation}
for $x_f= 25$ where $x_f=\Mdm/T_f$ and $T_f$ is the temperature at
freeze-out. The averaged velocities in Eq.~(\ref{eq:vfo}) represent a
mild suppression but which is enforced by the distinct $r =
M_{med}/\Mdm$ dependence of the 2-body cross sections,
\begin{equation}
\label{eq:SS}
  \langle \sigma v\rangle (S S \rightarrow l\bar l) \approx {\langle v^4\rangle}\,
{ y_S^4 \over 60 \pi}\, {M_S^6\over M_E^8}\nonumber
\end{equation}
(compare with Eqs.~(\ref{eq:NN}) and (\ref{eq:chichi})).  Hence, for fixed
$r$, DM mass and thermal velocity, it is clear that the coupling $y_S$
must be larger than $y_\chi$ or $g_N$ to match the observed relic
abundance. This, as shown in \cite{Toma:2013bka,Giacchino:2013bta},
has interesting implications for the strength of radiative processes.

\begin{figure}
 \includegraphics[width=9cm]{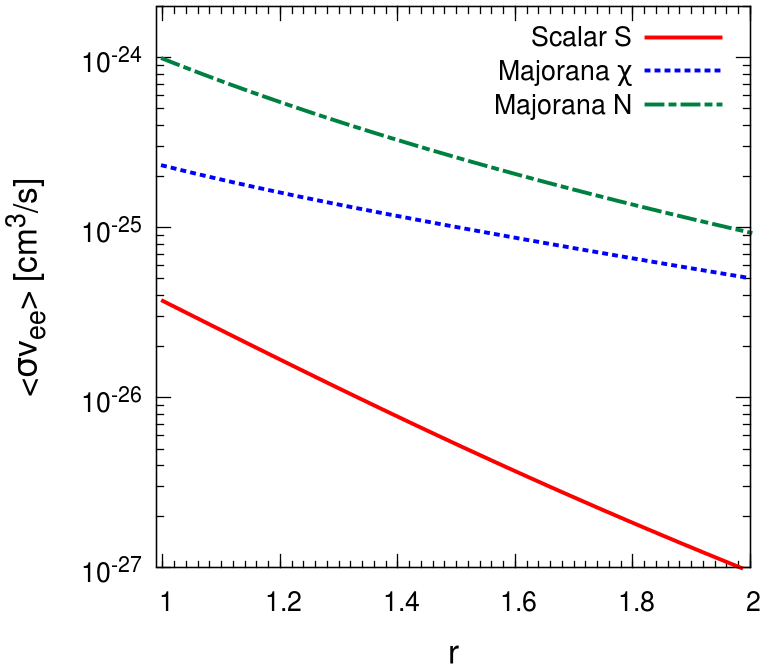}
  \caption{Annihilation cross sections into a SM fermion pair as a function of 
    $r =M_{med}/M_{DM}$ (thermal averages and in the chiral limit). The cross sections are given
    for unit couplings and for $M_{DM}=100$ GeV.}
  \label{fig:sv2all}
\end{figure}

\section{Spectral energy density: internal Bremsstrahlung} 
\label{sec:vib}

In this section we discuss the contribution of so-called internal
Bremsstrahlung to the spectral energy density of gamma rays. This has
been discussed extensively in the literature, so we just recap the
salient features.  Bremsstrahlung is of interest for two reasons.
First, the annihilation cross section in a s-wave through
Bremsstrahlung is no longer mass suppressed. For one thing, there is
no obstruction from conservation of angular momentum, but there is
more to it. Although the argument is somewhat gauge-dependent, this
result may be traced to emission of a gamma ray from the virtual
massive charged particle in the t- and u-channels, or so-called virtual
internal Bremsstrahlung (see {\em e.g.} \cite{Bringmann:2007nk}).
This implies that the Bremsstrahlung, a 3-body final state process,
may be more important than the 2-body tree level process, despite the
suppression of the former by a factor ${\cal O}(\alpha/\pi)$.  This is
typically the case for annihilation in light fermions or equivalently
heavy dark matter $\Mdm\gg m_f$, and when the velocity is
non-relativistic, like at the galactic center ($v \sim 10^{-3}$).
Second, emission from the virtual mediator, depending on the ratio $r
= M_{med}/\Mdm$, may have a sharp spectral feature, possibly
mimicking a monochromatic gamma ray line.

For reference, we give here the expressions of the 3-body annihilation
cross section for the 3 scenarios. Defining
\begin{eqnarray}
  vd\sigma_{2\rightarrow 3}&=&\frac{|\M|^2}{128\pi^3} dxdy
\label{eq:dsv3}
\end{eqnarray}
where $v={\sqrt{{k_1 \cdot k_2}-m_1m_2}}/{E_1E_2}$ refers to the
relative velocity of the $S$ particles and $x, y$ are the reduced
energy parameters $x=2E_\gamma/\sqrt{s}$ and $y=2E_f/\sqrt{s}$, with $s$
the Mandelstam variable corresponding to the center-of-mass energy
squared, we have:

\subsubsection{Scenario $1$: $\chi$ DM candidate \cite{Bergstrom:1989jr}}
\begin{equation}
  \label{eq:amp3bodyChi}
{1\over 4} \sum_{\rm spin}|\M_\chi|^2=\frac{4 \, \pi \,\alpha
  \,y_\chi^4}{M_\chi^2 } \, \frac{ 4 (1 - y) (2 + 2 x^2 + 2 x( y-2) - 2 y
  + y^2) }{(1 -r_\chi^2 - 2 x)^2 (3 +r_\chi^2 - 2 x - 2 y)^2}
\end{equation}
\subsubsection{Scenario $2$: $N$ DM candidate \cite{Barger:2011jg}}
\begin{equation}
  \label{eq:amp3bodyN}
{1\over 4} \sum_{\rm spin}|\M_N|^2=\frac{ \, \pi \,\alpha
  \,g_N^4}{M_N^2 }\left(2 + {1\over  r_N^2}\right)^2 \, \frac{ 4 (1 - y) (2 + 2 x^2 + 2 x( y-2) - 2 y
  + y^2) }{(1 -r_N^2 - 2 x)^2 (3 +r_N^2 - 2 x - 2 y)^2}
\end{equation}

\subsubsection{Scenario $3$: $S$ DM candidate \cite{Barger:2011jg,Toma:2013bka,Giacchino:2013bta}}
\begin{equation}
\label{eq:amp3bodyS}
 |\M_S|^2=\frac{32 \pi\, \alpha \,y_S^4 }{ M_S^2}\,\frac{ 4  (1 - y) (2 + 2 x^2 + 2 x ( y-2) - 2 y + y^2) }{
 (1 - r_S^2 - 2 x)^2 (3 + r_S^2- 2 x - 2 y)^2}  
\end{equation}

\bigskip

The acute reader will have noticed that, for fixed $r_{\chi,N,S}$, the
dependence of the Bremsstrahlung cross sections into gamma rays is
precisely the same in
Eqs.~(\ref{eq:amp3bodyChi},\ref{eq:amp3bodyN},\ref{eq:amp3bodyS}),
which implies that all three scenarios have the same spectral
signature \cite{Barger:2011jg}. Notice that we have already checked
this dependence for both scenarios $1$ and $3$
in~\cite{Giacchino:2013bta}. The Bremsstrahlung cross section in the
scalar case was also re-derived in~\cite{Toma:2013bka} and previously
obtained in the large mediator mass regime in~\cite{Boehm:2006df}.

\bigskip
An argument to explain this conclusion has been advanced in
\cite{Barger:2011jg} based on effective operators. First there should
be no distinction between scenarios $1$ and $2$ since they have
the same initial and final states. As above, the bi-linear operator
corresponding to the initial states is $\bar \chi \gamma_5 \chi$. Then
they have shown that Bremsstrahlung corresponds to the effective
operator
\begin{equation}
  {\cal O}_\chi \sim \bar\chi \gamma_5 \chi\, \left(\partial_\mu \bar \psi_R \gamma_\nu \psi_R + \bar \psi_R \gamma_\nu \partial_\mu \psi_R\right) \tilde F^{\mu\nu}\, ,
\label{eq:effOggC}
\end{equation}
 where $\tilde F^{\mu\nu}$ is the dual of $F^{\mu\nu}$.  In scenario
$3$, the initial state corresponds to $S^2$ and the effective
coupling is given by
\begin{equation}
   {\cal O}_S \sim S^2\, \left(\partial_\mu \bar \psi_R \gamma_\nu
\psi_R + \bar \psi_R \gamma_\nu \partial_\mu \psi_R\right) F^{\mu\nu}
\label{eq:effOggS}
\end{equation}
The only difference amounts to exchanging the role of the $\vec E$ and
the $\vec B$ of the photon and so the spectra are the same (up to
normalization). This argument is essentially based on a rephrasing of
the exact result in terms of effective operators, and thus is not {\em
  per se} an explanation, but it may complemented as follows. Notice
that the effective couplings correspond to respectively a dimension 9
and 8 operator, but there are other, {\em a priori} independent,
operators. Incidentally a classification of all dimension 8 operators
contributing to photon Bremsstrahlung for both Majorana and scalar DM
has been given in \cite{DeSimone:2013gj}: for Majorana DM there are 5
operators, out of which 3 are CP even, while in the scalar case, there
are 7 operators, with 4 being CP even. Remarkably, while these
operators lead to different spectra for the emission of $W$ and $Z$,
the spectra for emission of gamma rays are precisely the same (see
Eqs.~(3.8)-(3.10) and Eqs.~(3.24)-(3.27) in
\cite{DeSimone:2013gj}). Moreover, it can be easily checked that they
have the same $x$ and $y$ dependence as the numerator of
Eqs.~(\ref{eq:amp3bodyChi}),~(\ref{eq:amp3bodyN}) and
(\ref{eq:amp3bodyS}) (after some obvious change of variables). This, in
passing, means that the dimension 9 operator for Majorana DM of
Eq.~(\ref{eq:effOggC}) must be equivalent to (a combination of) the
dimension 8 operators studied in \cite{DeSimone:2013gj}.  While the
$r$ dependence of the denominator in
Eqs.~(\ref{eq:amp3bodyChi})-(\ref{eq:amp3bodyS}) can not be obtained from
an effective approach, it may be easily inferred from the propagators
of the intermediate particles. Hence the effective operator argument that Bremsstrahlung
from scalar and Majorana DM should have the same spectra is actually robust.

\bigskip

\begin{figure}
 \includegraphics[width=9cm]{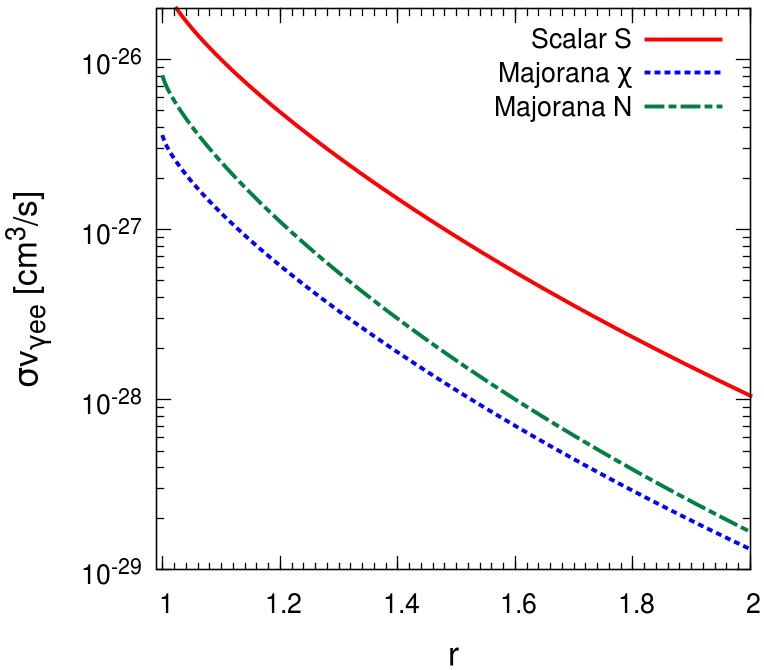}
  \caption{Total 3-body cross sections (gamma ray Bremsstrahlung emision). Same conventions as in Fig.~\ref{fig:sv2all}.}
  \label{fig:sv3all}
\end{figure} 

Now it remains that the normalizations of the spectra are distinct, or
at least they are for the scalar, for scenarios $1$ and $2$ are quite
similar. For equal $\Mdm$, couplings and $r$, the amplitude in the
gauge case is larger by a factor of
$$
{1\over 2}\left(2+ {M_N^2\over M_{W^\prime}^2}\right) \sim 1
$$ 
where the first term is from the 2 transverse polarization modes
and the second term from the longitudinal one.  As emphasized in
\cite{Toma:2013bka,Giacchino:2013bta}, all things being taken to be
the same ({\em i.e.} DM and mediator masses and the couplings), the 3-body cross section is larger by a factor of 8 in
scenario $3$ compared to scenario $1$. This, together with the
relative suppression of the 2-body cross section leads to an enhanced
gamma ray feature in the scalar case compare to the Majorana cases.

\bigskip
The intermediate conclusion is that scenarios $1$ and $2$ are
essentially identical. They share the same spectra, with quasi the
same parametric dependence in the couplings and mass of DM and of the
mediator of the normalization of the 2 and 3-body processes. The
scalar case is distinct, in the sense that if the relic abundance is
thermal and is fixed by the 2-body annihilation process, then the
signal is stronger for the scalar case (by a factor which may be as
large as 2 orders of magnitude)
\cite{Toma:2013bka,Giacchino:2013bta}. If we relax the latter
constraint (for instance if the abundance is fixed by another
process), then the 3 scenarios become indistinguishable. It is thus of
interest to check whether other spectral features may help to lift
this degeneracy. The spectrum of gamma rays from say, $\pi_0$, being
featureless we focus on gamma ray lines. In \cite{Giacchino:2013bta}
we have tentatively included the features from annihilation into two
monochromatic gamma rays. However the one-loop cross sections, in
particular that relevant for scenario $3$, reported in the
literature has some peculiarities. We have thus felt compelled to
reanalyze this problem. Our results are presented in the next section.

\begin{figure}
 \includegraphics[width=9cm]{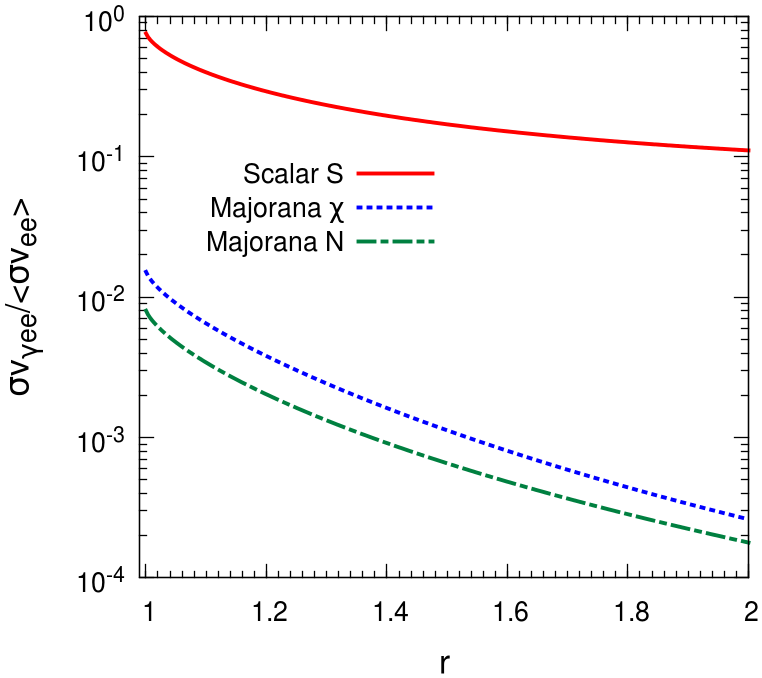}
  \caption{Ratios of the 3-body (gamma ray Bremsstrahlung emision) to 2-body annihilation cross sections. The 2-body cross section is thermally averaged over velocities at the time of freeze-out, relevant to determine the relic abundance of DM.  }
  \label{fig:sv2gllallM}
\end{figure}
\section{SPECTRAL ENERGY DENSITY: GAMMA RAY LINES}
\label{sec:annih-into-gamma}

We consider the annihilation of non-relativistic DM into two on-shell gamma rays,
$$
DM(p_1)  + DM(p_2) \rightarrow \gamma(\epsilon_1,k_1) + \gamma(\epsilon_2,k_2)
$$ with $p_{1, 2}$ and $k_{1,2}$ the momenta, $\epsilon_{1,2}$ the
polarization vectors, $p_1 + p_2 = k_1 + k_2$, and $\epsilon_1 \cdot
k_1 = \epsilon_2\cdot k_2 = 0$.

 In the 3 scenarios we consider the amplitude for this process is
 represented by a sum of box Feynman diagrams. Although the individual
 diagrams may be infinite, the total amplitudes are finite and are
 moreover non-vanishing in the s-wave. These features allow
 substantial simplifications in the calculation of the Feynman
 diagrams. In particular we can calculate the amplitude in the limit
 $p_1 = p_2 = (\Mdm,\vec 0)$. In this limit, Gram determinants built on
 the external momenta or their relevant linear combinations may be
 vanishing. For instance
$$
\left\vert\begin{array}{ccc}
p_1^2 & p_1 \cdot p_2 & p_1 \cdot k_1\\
p_1\cdot p_2 & p_2^2 & p_2 \cdot k_1\\
p_1\cdot k_1 & p_2 \cdot k_1 & k_1 \cdot k_1\\
\end{array}
\right\vert = 0 
$$ The standard Passarino-Veltman reduction of tensorial loop
integrals breaks down when Gram determinant are zero
\cite{Passarino:1978jh}. This is in particular an issue\footnote{To be fair we should mentioned that we have managed to obtain
  numerical results from FormCalc that are in very good agreement with our
  analytical expressions. } for automated
tools like FormCalc and is also a source of numerical
instabilities in LoopTools
\cite{Hahn:1998yk}.  Fortunately, by the very same token, one may use
the degeneracy between the momenta to express 4-point integrals in
terms of 3-point integrals, and also some 3-points integrals in terms
of 2-points integrals \cite{Stuart:1987tt} (see also
\cite{Bergstrom:1997fh,Bertone:2009cb}). In particular this implies
that no 4-points loop integrals appear in the final expression of the
amplitude, which may be expressed in terms of finite, 3-points loop
integrals that are much easier to handle, and in particular may be
given in terms of rather simple analytic functions. As in the previous
section, we discuss the 3 scenarios separately.

\subsubsection{Scenario $1$: $\chi \chi \rightarrow \gamma \gamma$}

This process has been calculated several times, starting from the
seminal works of \cite{Bergstrom:1988fp,Rudaz:1989ij,Bergstrom:1989jr}
in the context of supersymmetry. As is well known, in the limit
$M_{\tilde E} \gg M_\chi$, the amplitude for $\chi \chi \rightarrow
\gamma \gamma$ may be related to the chiral anomaly. A remarkable
consequence is that the amplitude does not vanish in the chiral limit,
$m_f \rightarrow 0$.

For our own sake and to check our procedures, we have
redone the calculation of this amplitude in the non-relativistic
limit, $v_\chi \rightarrow 0$, albeit with the help of FeynCalc
\cite{Mertig:1990an}. To do so we have used the trick of
\cite{Bergstrom:1997fh} which consists of projecting the initial
$\chi$ pair into a s-wave state, using
$$
{\cal O} = -{M_\chi \over \sqrt{2}}\gamma_5 \left(1 - \gamma^0\right)\,.
$$ 
In the chiral limit the amplitude involves a single 3-points scalar
loop integral.  For reference, we give its expression following the
standard nomenclature of 3-point loop integrals (the $C_0$ functions of
Passarino and Veltman) and then explicitly in terms of analytic
functions.  We have found
\begin{eqnarray}
\langle\sigma v\rangle_{\gamma\gamma}^\chi=\frac{y_\chi^4\alpha^2M_\chi^2}{64\pi^3}  \left|C_0\left(-M_\chi^2,M_\chi^2 , 0, r_\chi^2 M_\chi^2,0, r_\chi^2M_\chi^2\right)\right|^2\,.
\end{eqnarray}
Using
\begin{eqnarray}
\label{eq:C0_1}
C_0(- M^2,M^2,0,r^2 M^2,0,r^2 M^2)  &=& {-1\over 2 M^2} \int_0^1 {dx\over x} \log\left(\left\vert{-x^2 + (1-r^2)x + r^2\over x^2 - (1+r^2) x+ r^2}\right\vert \right)\\
&=& {-1\over 2 M^2}\left(Li_2\left({1\over r^2}\right) -  Li_2\left({-1\over r^2}\right)   \right)
\end{eqnarray}
we are in complete agreement with previous results and in particular
with \cite{Bergstrom:1997fh,Bern:1997ng} (we have also calculated the amplitude
for the case $m_f \neq 0$ - the result may be read from
\cite{Bergstrom:1997fh}). Of interest for us will be the following
limits,
\begin{eqnarray}
\svgg^\chi&=&\frac{y_\chi^4\alpha^2\pi}{64^2 M_\chi^2} \quad \mbox{for } r_\chi=1 \quad\mbox{and }\quad
\svgg^\chi\approx\frac{y_\chi^4\alpha^2}{64 \pi^3}{M_\chi^2\over M_{\tilde E}^4} \quad \mbox{for } r_\chi\gg 1
\end{eqnarray}

\subsubsection{Scenario $2$: $ N N  \rightarrow \gamma \gamma$}

For this process we refer to the work of \cite{Bergstrom:1997fh} in
which a related amplitude has been calculated. Specifically, the
process considered there is the annihilation of two neutralinos into
two photons through a chargino and SM $W^\pm$ gauge boson
loops. Making simple adjustments in the couplings and the particle
content, and taking into account the coupling to nonphysical Goldstone
modes (as alluded to above, the calculation of \cite{Bergstrom:1997fh}
has been done in a 't~Hooft-Feynman non-linear $R_\xi$ gauge), we get
\begin{eqnarray}
\label{eq:NNgg}
\svgg^{N}& =&\frac{g_N^4\alpha^2 M_N^2}{64 \pi^3}\left|4 C_0(4 M_N^2,0,0,r_N^2 M_N^2,r_N^2 M_N^2,r_N^2 M_N^2)\right.\nonumber \\
&& - \left.\left(2+\frac{1}{r_N^2}\right) C_0(-M_N^2,M_N^2,0, r_N^2 M_N^2, 0, r_N^2 M_N^2)\right|^2
\end{eqnarray}
with 
\begin{eqnarray}
\label{eq:C0_2}
C_0(4 M^2,0,0,r^2 M^2,r^2 M^2,r^2 M^2) &=&{1\over 4 M^2} \int_0^1\frac{dx}{x}\log\left(\left|4\frac{x^2}{r^2}-4\frac{x}{r^2}-1\right|\right)\cr
 &=& {-1\over 2 M^2} \left( \arctan{1\over \sqrt{r^2-1}}\right)^2
\end{eqnarray}
where the last equality holds for $r\geq 1$ and the other $C_0$ are as
defined in Eq.~(\ref{eq:C0_1}). The limiting behaviours are now given
by
\begin{eqnarray}
\svgg^N&=&\frac{g_N^4\alpha^2\pi}{64^2 M_N^2}\frac{25}{4} \quad \mbox{for } r_N=1 \quad\mbox{and }\quad
\svgg^N\approx\frac{g_N^4\alpha^2}{64 \pi^3}{M_N^2\over M_{W^\prime}^4} \quad \mbox{for } r_N\gg 1\,.
\label{eq:limN}
\end{eqnarray}

As an independent check of this result, we have verified that we can
recover the cross-sections derived many years ago by Crewther {\em et
  al} \cite{Crewther:1981wh}, for the annihilation of SM neutrinos into
two photons, in a regime corresponding to our limit $r_N\gg 1$ in
Eq.~(\ref{eq:limN}) (see Appendix~\ref{sec:appendixA}).

\subsubsection{Scenario $3$: $ S S  \rightarrow \gamma \gamma$}
\label{sec:Sgg}

For this case we have redone the full one-loop calculation, including
finite SM fermion mass $m_f \neq 0$ contributions. The full expression
is given in the Appendix~\ref{sec:appendix-B}. Here we just give the expression in the
chiral limit, as for the other two scenarios.

To derive the amplitude, we have essentially followed the path of
\cite{Bertone:2009cb}. We made use of FeynCalc and calculated
amplitude and cross section for S particles at rest ($v_S = 0$).
This amounts to evaluate a combination of one-loop 4-points tensor
integrals. Since the Gram determinant is zero for DM particles at
rest, a straightforward application of Passarino-Veltman reduction
does not work, so instead we used the approach of \cite{Stuart:1987tt}
to express directly the 4-points loop integrals as a linear
combination of 3-points scalar integrals.

Writing
\begin{equation}
  \svgg = \frac{2\,y_S^4\alpha^2}{64\pi^3M_S^2}|{\cal A}|^2
\label{eq:sv}
\end{equation}
at an intermediate step we got the following expression (for $m_f =0$)
\begin{eqnarray}
{\cal A}&=& 2 +\frac{2}{(1 - r_S^2)}B_0\left(M_S^2, 0, r_S^2M_S^2\right) - B_0\left(4 M_S^2, 0, 0\right) 
- \frac{1 + r_S^2}{1 - r_S^2}B_0\left(4 M_S^2, r^2M_S^2, r_S^2M_S^2\right)\nonumber\\
&& + M_S^2\left(
- (1 + r_S^2)\left(C_0\left(M_S^2, M_S^2, 4 M_S^2, 0, r_S^2M_S^2, 0\right)
+C_0\left(M_S^2, M_S^2, 4 M_S^2, r^2M_S^2, 0, r_S^2 M_S^2\right)\right) \right.\nonumber\\
&&\left.
- 2C_0\left(-M_S^2, M_S^2, 0, r_S^2 M_S^2, 0, r_S^2 M_S^2\right)
+ 4r_S^2C_0\left(4M_S^2, 0, 0, r_S^2 M_S^2, r_S^2 M_S^2, r_S^2 M_S^2\right)\right)\,.\nonumber\\
 \label{eq:Sgg-us}
\end{eqnarray}
 Despite the presence of divergent 2-point integrals, this expression
 is finite. Actually it may be simplified using the fact that some
 3-point scalars integral, whose momentum arguments have a vanishing
 Gram determinant, may be reduced further
\begin{eqnarray}
  C_0\left(M_S^2, M_S^2, 4M_S^2, M_E^2, 0, M_E^2\right) &=& \frac{B_0\left(M_S^2, 0, M_E^2\right)- B_0\left(4M_S^2, M_E^2, M_E^2\right)}{ - M_E^2 + M_S^2}\\ 
  C_0\left(M_S^2, M_S^2, 4M_S^2, 0, M_E^2, 0\right) &=& \frac{B_0\left(M_S^2, 0, M_E^2\right) - B_0\left(4M_S^2, 0, 0\right)}{ M_E^2 + M_S^2}\label{eq:subst}
\end{eqnarray}
Using this substitution, we get the following simple expression
\begin{equation}
{\cal A}= 2 - 2 M_S^2 C_0(-M_S^2, M_S^2, 0, r_S^2 M_S^2, 0, r_S^2 M_S^2) + 4 r_S^2 M_S^2C_0(4 Ms^2, 0, 0,r_S^2 M_S^2, r_S^2M_S^2, r_S^2M_S^2)\,.
\label{eq:Sgg-uss}
\end{equation}
where the two 3-points loop integrals are those given in
Eqs.~(\ref{eq:C0_1},\ref{eq:C0_2}).\footnote{Notice that our result differs from that reported in Eq.~(13) of \cite{Bertone:2009cb}. We believe that the discrepancy is just a mere misprint in the last line of their Eq.~(13), in which $C_0\left(M_S^2, 0, M_S^2, 0, r_S^2 M_S^2, r_S^2 M_S^2\right)$ should read $C_0\left(-M_S^2, 0, M_S^2, 0,r_S^2M_S^2, r_S^2M_S^2\right)$.  Although virtually impossible to spot,  this becomes pretty clear when one realizes that no combination of external momenta of the box Feynman diagrams may lead to the arguments of $C_0\left(M_S^2, 0, M_S^2, 0, r_S^2 M_S^2, 0, r_S^2 M_S^2\right)$. The erroneous expression is divergent at $r=1$, and thus potentially leads to a large signal for annihilation into gamma rays ~\cite{Tulin:2012uq}. On the contrary we found that the correct expression is regular at $r=1$.\label{fn:typo}}

Finally for  $r_S =1$ and $r_S \gg 1$, we have respectively
\begin{eqnarray}
\svgg^S&=& \frac{y_S^4\alpha^2}{8 \pi^3 M_S^2}\left( 1-\frac{\pi^2}{8}\right)^2\quad \quad\mbox{and }\quad
\svgg^S\approx\frac{y_S^4\alpha^2}{18\pi^3 M_S^2 r^4} \hspace*{1cm}
\label{eq:SSgglimits}
\end{eqnarray}

\begin{figure}
 \includegraphics[width=9cm]{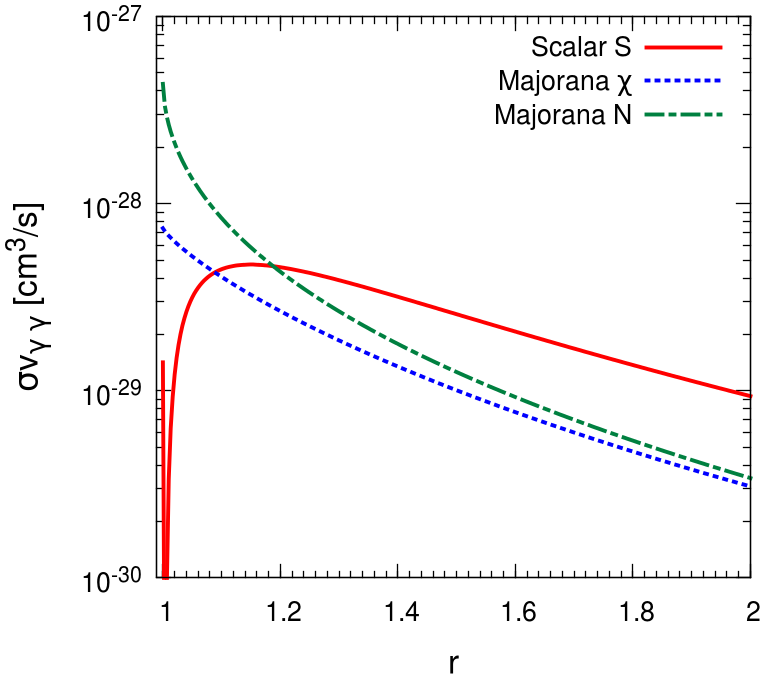}
  \caption{Annihilation into 2 photons. Same conventions as in Fig.\ref{fig:sv2all}.}
  \label{fig:svggall}
\end{figure}

\begin{figure}
 \includegraphics[width=9cm]{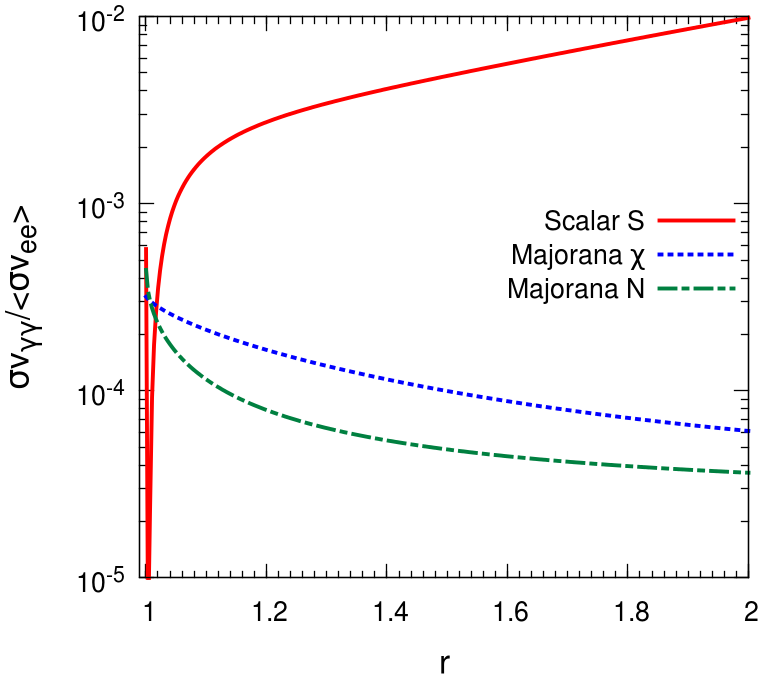}
  \caption{Ratios of the cross sections into 2 photons to the 2-body annihilation into a fermion pair.}
\label{fig:sv2ggallM}
\end{figure}


\section{Discussion of results}

In this section we compare the salient features of the 3 scenarios
considered in the previous section. The thermally averaged 2-body
annihilations cross sections into two fermions at the time of
freeze-out (i.e. for averaged velocities taken as in
Eq.~(\ref{eq:vfo})) in the chiral limit are shown in
Fig.~\ref{fig:sv2all}. For convenience we have normalized all the
couplings to 1 and we took $\Mdm=$ 100 GeV. This figure shows that the
cross section in the scalar DM scenario (scenario $3$) is
parametrically smaller than that of scenarios with Majorana DM
(scenarios 1 and 2): while scenarios 1 and 2 (for $\chi$ and $N$ DM) share
the same asymptotic behaviour for large $r$, the thermally averaged
cross section in scenario 3 (for $S$ DM) for, say $r=2$, is suppressed by
almost 2 orders of magnitude. This result is due to a combination of
the d-wave dependence of the cross section and of a distinct
dependence on $r$, see Eqs.~(\ref{eq:chichi}), (\ref{eq:NN})
and~(\ref{eq:SS}). If the relic abundance is fixed by the 2-body
process, a larger coupling is thus required for the $S$ than for the
$\chi$ or $N$ scenarios \cite{Toma:2013bka,Giacchino:2013bta}.

In Fig.~\ref{fig:sv3all}, we give the 3-body annihilation cross
section (for photon Bremsstrahlung emission) again for unit couplings
and $M_{DM} = 100$ GeV. The $r$-dependences are very similar, but the
signal from $S$ is comparatively larger compared to the $\chi$ and $N$
scenarios, respectively by a factor $8$ and $8/(1+1/2 r^2)^2$. This,
combined with the previous feature, implies that the Bremsstrahlung is
potentially much stronger in the $S$ scenario, although the spectra
are the same \cite{Toma:2013bka,Giacchino:2013bta}. This is
illustrated in Fig.~\ref{fig:sv2gllallM} which shows the ratio of the
3-body to 2-body cross sections (note that this ratio is independent
of the couplings and of the mass of the DM candidate). If the 2-body
annihilation into two fermions is the one driving the relic abundance
in the early universe, the Bremsstrahlung spectral feature is expected
to be much stronger in the scalar DM scenario ($\svgll$ is almost 3
orders of magnitude larger than for Majorana DM for $r\sim 2$).

\begin{figure}
 \includegraphics[width=9cm]{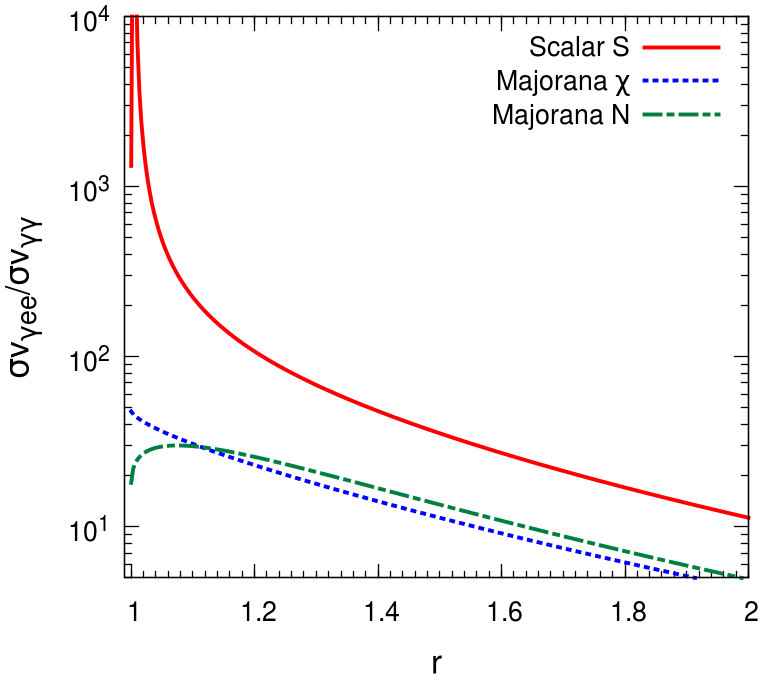}
  \caption{Ratios of the Bremsstrahlung to 2 photons cross sections.}
\label{fig:sv3ggall}
\end{figure}

Turning to the annihilation into monochromatic gamma rays, we compare
the three cross sections in Fig.~\ref{fig:svggall}. The dependence in
the $\chi$ scenario is well-known, and has been reported many times in
the literature. In particular, it may be related to the chiral
anomaly, which implies that the cross section is non-vanishing even in
the chiral limit \cite{Rudaz:1989ij,Bergstrom:1989jr}. Since the
initial and final states are the same in scenarios $1$ and $2$, one
may expect a similar behaviour, which is confirmed by the calculations
and is illustrated in Fig.~\ref{fig:svggall}. Although this result is
implicit in the literature (we derive the $N$ cross section from the
results on the neutralino discussed in \cite{Bergstrom:1997fh}), we
are not aware of an explicit discussion in the framework of simpler
models, like that of \cite{Khalil:2009nb,Ma:2009tc}. At any rate, the
cross section is that given in Eq.~(\ref{eq:NNgg}).  It is slightly
larger than in the $\chi$ scenario for $r$ close to 1 (by a factor of
$25/4$ at $r=1$), but asymptots to the same result at large $r$. The
large $r$ behaviour is also in agreement with the result of Crewther
{\em et al} for the annihilation of Dirac neutrinos into two photons
\cite{Crewther:1981wh}. Incidentally, the latter result has been
derived making use of the chiral anomaly, as in the early derivation
of the cross section in the $\chi$ scenario.

The behaviour of the cross section in scenario $3$ is more
puzzling. Having done the calculation using different approaches, and
also having obtained the same results as those reported in
\cite{Ibarraetal}, we are confident that the expression is correct. In
particular it is finite at $r=1$, albeit with a strange value compared
to the Majorana cases. The large $r$ behaviour is also completely
mundane, having the same dependence in $r$ as in scenarios $1$ and
$2$. It shows however peculiar feature, since it has a maximum around
$r=1.15$ and then a dip, with a zero near $r=0$. The origin of
this destructive interference is unclear, at least to us, although it
may be traced to amplitudes that correspond to Feynman diagrams with a
distinct number of heavy fermion propagators. It may be of academic
interest to investigate this phenomenon further, which may perhaps be
related to the distinct property of the tree level annihilation cross
section into fermion pairs in the $S$ scenario compared to the
Majorana cases. Another, perhaps not unrelated question is whether the
annihilation cross section of $S$ into two photons may be derived from
the trace anomaly. 

This being said, we see that the $S$ cross section is larger by a
factor of $32/9 \approx 3.5$ than in the Majorana DM scenarios, which
have the same asymptotic behaviour. The annihilation of a scalar may
thus also lead to a relatively stronger signal into monochromatic
photons. This is shown in Fig.~\ref{fig:sv2ggallM}, which displays
the ratio of the cross section into 2 gamma rays to that into a
fermion pair. The rise as a function of $r$ of the signal in the $S$
case is due to the fact that the 2-body cross section into fermion
pairs scales like $r^{-6}$, see Eq.~(\ref{eq:SS}), while the
annihilation into 2 photons is $\propto r^{-4}$, see
Eq.~(\ref{eq:SSgglimits}). On the contrary the ratios asymptote to a
small constant, ${\cal O}(10^{-5})$, for the Majorana scenarios.
\begin{figure}
 \includegraphics[width=9cm]{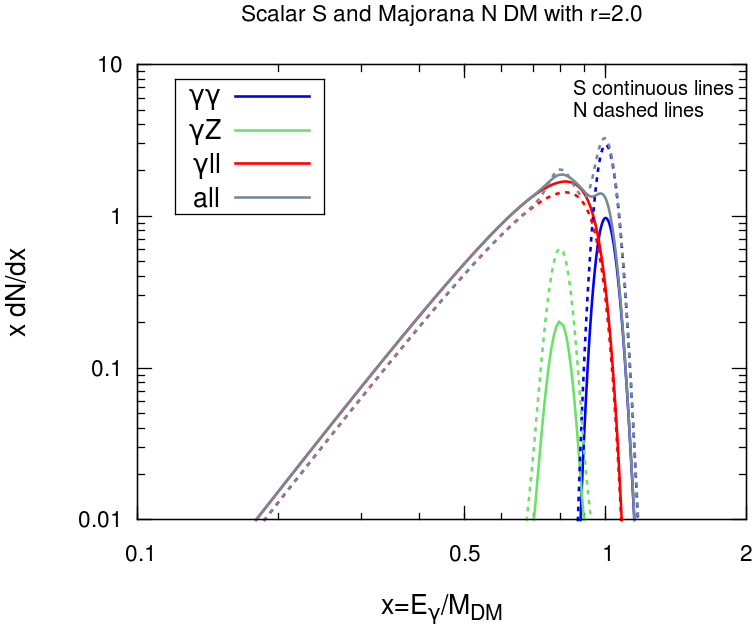}
  \caption{Comparison of normalized photon spectra for scenarios 2 and 3 for $\Mdm = 100$ GeV.}
\label{fig:LSpectra-m100-r20-SC}
\end{figure}

For completeness we also give in Fig.~\ref{fig:sv3ggall} the ratio of
3-body annihilation cross section (photon Bremsstrahlung emission) to
the one into two photons.  The Bremsstrahlung signal becomes
relatively less prominent for large $r$, as the cross sections drop
like $r^{-6}$, but we see it is more dominant in the scalar than in
the Majorana scenarios. Hence both signals are stronger for the
scalar. The relative importance of the $\gamma \gamma$ signal compared
to the $\gamma ee$ one is however comparatively smaller in the scalar
case than in the Majorana DM cases.  This may be seen directly in the
photon spectra of Fig.~\ref{fig:LSpectra-m100-r20-SC} where we compare
scenario 2 and 3 ($N$ and $S$) for $r=2$; scenario 1 ($\chi$) would
have essentially the same signature as scenario 2. The quantity $x dN/dx=
E_\gamma/\sigma v_{\gamma} d\sigma v_{\gamma i}/dE_\gamma$, with
$i=\gamma, ee$ and $\sigma v_{\gamma} = \sum_i\sigma v_{\gamma i}$,
denotes the normalized photon spectrum multiplied by the photon
energy. We see that for $r=2$ the dominant Bremsstrahlung feature gets
an extra contribution from the $\gamma \gamma$ and $\gamma Z$
lines. Notice that we have assumed an energy resolution of $\Delta
E/E=0.1$, and that, although we did not explicitly derived $\svgZ$
here, we follow the same procedure as in~\cite{Giacchino:2013bta} to
estimate the $\gamma Z$ contribution. We refer to
\cite{Toma:2013bka,Giacchino:2013bta} for some discussion of scenarios
$1$ and $3$. Clearly scenario $2$ should give a phenomenology similar
to that of the $\chi$ Majorana. Here we just provide a few benchmark
values (see Table \ref{tab:m100}).

\begin{table}[t]
  \begin{tabular}{c|cc|cc|cc}
     &\multicolumn{2}{c|}{Scenario $1$ ($\chi$)} &\multicolumn{2}{c|}{Scenario 2 ($N$)}&\multicolumn{2}{c}{Scenario 3 ($S$)} \\
    \hline
    &$r=$1.2&$r=2$&$r=1.2$&$r=2$&$r=1.2$&$r=2$\\
    \hline
$\svgll$&$6.1\, 10^{-28}$&$1.3\, 10^{-29}$&$1.1\, 10^{-27}$&$1.7\, 10^{-29}$&$4.9\, 10^{-27}$&$1.1\, 10^{-28}$ \\
$\svgg$&$2.9\, 10^{-29}$&$3.1\,10^{-30}$& $4.3\, 10^{-29}$&$ 3.4\, 10^{-30}$&$4.6\, 10^{-29}$&$9.4\,10^{-30}$\\
\hline
  \end{tabular}
  \caption{Cross sections in units of cm$^3$/s for $\Mdm=100$ GeV and
    unit couplings}
  \label{tab:m100}
\end{table}
\section{Conclusions}

In this work we have complemented the work of Barger {\em et al}, in
which 3 simple scenarios of DM with Bremsstrahlung of photons are
discussed. It also complements the phenomenological studies initiated
in \cite{Toma:2013bka} and \cite{Giacchino:2013bta} in which it has
been shown that a real scalar DM candidate $S$ interacting with light
SM fermions could give a strong Bremsstrahlung signal. Specifically we
have considered the radiative annihilation of three DM candidates into
two photons. This lead us to re-calculate the amplitude for $SS
\rightarrow \gamma \gamma$. Our result differs from expressions that
may be found in the literature, but the discrepancy is minor, being
likely due to a misprint, which however is difficult to spot without
actually doing the full calculation.  Hence we believe that it was
useful to provide an independent check of the expression for $SS
\rightarrow \gamma \gamma$. We have compared the result to those
expected in the case of a Majorana DM, interacting with SM fermions
either through a heavy charged scalar particle (scenario 1) or through
a charged gauge boson (scenario 2). The main outcome, which complement
the conclusions drawn in \cite{Toma:2013bka} and
\cite{Giacchino:2013bta}, is that radiative processes are
significantly more relevant for the scenario with scalar DM than for
the Majorana cases. These results from a combination of factors. First
the fact that the annihilation cross section into fermion pairs is
d-wave suppressed in the chiral limit in the scalar DM case, and
second, the fact that the radiative cross section are parametrically
larger. These results may be directly inferred from the analytical
expressions given in the body of the paper, and from a glance at the
figures.

\appendix
\section{From $\bar NN\rightarrow \gamma\gamma$ to $\bar\nu\nu\rightarrow \gamma\gamma$}
\label{sec:appendixA}

In this appendix we compare the cross section for annihilation in two photons in scenario 2 to the annihilation of Dirac SM neutrinos calculated by Crewther {\em et al} \cite{Crewther:1981wh}.
In~\cite{Crewther:1981wh}, they effectively made use of
the following Lagrangian:
\begin{equation}
{\cal L}_{eff}=\frac{G_F}{2\sqrt{2}}  \bar\nu(1-\gamma_5) \nu \,\bar e \gamma_\mu (1+4s_W^2-\gamma_5)e\, .
\label{eq:LeffC}
\end{equation}
They indeed show that all other possible contributions are
negligible. This has to be compared to the corresponding effective
Lagrangian resulting from our Eq.~(\ref{eq:LN}):
\begin{equation}
{\cal L}_{eff}=\frac{g_N^2}{8 M_{W'}^2}  \bar N(1+\gamma_5) N \,\bar e \gamma_\mu (1+\gamma_5)e\, .
\label{eq:Leffus}
\end{equation}
The main differences come from the overall factor: $\frac{g_N^2}{8
  M_{W'}^2}\rightarrow \frac{G_F}{2\sqrt{2}}$ and the vector
contributions to the neutrinos $N,\nu$ and the electron currents.  It
can be shown that none of the latter contributions to the $N,\nu,e$
currents are relevant for the final result.  In addition, Crewther
{\em et al} assumed that the SM neutrinos $\nu$ were of Dirac type
while in our case $N$ are Majorana neutrinos. This implies that our
final cross-section should be divided by a factor of 4 due to the fact
that two directions of the fermionic current flow are possible for the
Majorana particles to describe the same interaction (see also
e.g.~\cite{Rudaz:1989ij}). Departing from (\ref{eq:limN}) in the limit
$r_N\gg1$, we obtain for $M_N\rightarrow m_\nu$:
\begin{equation}
\svgg^\nu= \frac14\frac{\alpha^2 G_F^2 m_\nu^2}{8 \pi^3}\,,
 \label{eq:svggLMW}
\end{equation}
where the $\frac14$ factor comes from the Majorana to Dirac neutrino
change.  This equations has to be compared to equation (2.30) in
\cite{Crewther:1981wh} where their $\sigma c$ corresponds $\sigma
\sqrt{1-4m_\nu^2/s}= \svgg^\nu/2$ or equivalently to the cross-section
times the velocity of one dark matter
particle\footnote{$v/2=v_1=\sqrt{1-4m_\nu^2/s}$}. One can check that
our Eq.~(\ref{eq:svggLMW}) perfectly match their
Eqs. (2.28)-(2.30), realizing that in the chiral limit, their function
$I(m_e^2/s)\rightarrow 12 m_e^2/s$ and allows to cancel the $m_e$
insertions in their Eq.~(2.28).

\section{Full expression of $\svgg^S$}
\label{sec:appendix-B}

If $m_l\neq0$ the expression of the amplitude for $SS\rightarrow
\gamma \gamma$ (s-wave contribution) in Eq.~(\ref{eq:Sgg-uss}) should
be replaced by
\begin{eqnarray}
{\cal A}&=& 2+\left[ 4\,\frac{ m_l^2 (m_l^2 - M_S^2) C_0(4 M_S^2, 0, 0, m_l^2, m_l^2, m_l^2)}{m_l^2 - M_S^2(r_S^2+1)}\right. \nonumber\\
&&\hspace{1cm}+2 m_l^2 M_S^2\,  \frac{ \left(m_l^2(-m_l^2+2M_S^2)  +M_S^4(r_S^2  -1)\right) C_0(-M_S^2, M_S^2, 0, m_l^2, r_S^2 M_S^2, m_l^2)}{(m_l^2 - r_S^2 M_S^2) (m_l^2 - M_S^2(r_S^2+1)  ) 
   (m_l^2 - M_S^2(r_S^2-1))}\nonumber\\
&&\hspace{1cm}+(m_l\leftrightarrow r_S M_S)\bigg]
 \label{eq:Sgg-full}
\end{eqnarray}
 Notice that we have made use of (\ref{eq:subst})  to
obtain~(\ref{eq:Sgg-full}).

\section*{Note Added}
During the completion of this work we learned about the analysis
of~\cite{Ibarraetal} on the scalar dark matter scenario that includes
a gamma-ray spectral feature analysis. Their results agree with ours
in the aspects where our analysis overlap.

\acknowledgments

We thank Alejandro Ibarra, Takashi Toma, Maximilian Totzauer and
Sebastian Wild, who were working on a very similar project
(see~\cite{Ibarraetal}), for discussions and for sharing their results
with us.  We also thank C\'eline Boehm, Guillaume Drieu La Rochelle
and Maxim Pospelov for useful discussions. The work of F.G. and
M.T. is supported by the IISN, an ULB-ARC grant. LLH is supported
through an ``FWO-Vlaanderen'' post doctoral fellowship project number
1271513. LLH also recognizes partial support from the Strategic
Research Program ``High Energy Physics'' of the Vrije Universiteit
Brussel. All the authors are partly supported by the Belgian Federal
Science Policy through the Interuniversity Attraction Pole P7/37
``Fundamental Interactions''. M.T. also acknowledges support and
hospitality from the LPT at Universit\'e Paris-Sud and LLH
acknowledges hospitality and support from Nordita Institute at the
final stage of this work.

  \bibliography{bib3bdyM}{}

\end{document}